\documentclass[amsmath,amssymb,prc,final,showpacs,twocolumn]{revtex4-1} 

\usepackage{bm,hyphenat,xspace} 
\usepackage{graphicx,epsfig}

\newcommand{\scl}{0.62}

\newcommand{\A}[2]{{}^{#1}\mathrm{#2}}

\newcommand{\clh}{\mathcal{H}}

\begin{document}

\title {Core excitation in three-body nuclear reactions: 
Improved nucleon-core potential}

\author{A.~Deltuva} 
\email{arnoldas.deltuva@tfai.vu.lt}
\affiliation{Institute of Theoretical Physics and Astronomy, 
Vilnius University, A. Go\v{s}tauto 12, LT-01108 Vilnius, Lithuania}

\received{December 22, 2014}


\begin{abstract}
Three-body nuclear reactions in two-nucleon plus core systems are
described in the framework of exact scattering equations including
the core excitation. A nucleon-core optical potential is constructed
that can be easily adjusted to the reference potential
and thereby to the experimental two-body data, if available.
This constitutes an important improvement
over the simple deformation of the potential used previously
that violated the original fit to the data.
Predictions for  elastic, inelastic, and transfer reactions involving
$\A{10}{Be}$ and  $\A{24}{Mg}$ nuclear cores are obtained.
The new optical potential leads to a moderate increase of  cross sections.
\end{abstract}

 \maketitle

\section{Introduction \label{sec:intro}}

Deuteron ($d$) scattering from a nucleus $A$, consisting of $A$ nucleons, 
microscopically is an $A+2$ body  problem. 
While various $A\ge 3$ cases have been considered using approximate methods
\cite{quaglioni:08a}, a rigorous
solution of exact scattering equations \cite{faddeev:60a,alt:67a,yakubovsky:67,grassberger:67}
 has been achieved so far for three- and four-body systems only 
\cite{kievsky:01a,kuros:02b,deltuva:05c,deltuva:07d,lazauskas:12a,viviani:13a,deltuva:14a}. 
In addition, there exist also numerous approximate treatments
for three-body reactions \cite{johnson:70a,austern:87,baye:09a}. 
To apply the available three-body techniques to deuteron-nucleus scattering, 
this process  is often approximated by a 
three-body problem where the nuclear core $A$ is treated as a structureless particle whose
interaction with the proton ($p$) and the neutron ($n$) is given by complex or real potentials.
In some cases this may be a reasonable approach, but in others it is necessary to go beyond the
picture of an inert core $A$ and take into account its internal degrees of freedom.
It has been shown in several works \cite{crespo:11a,moro:12a,moro:12b,deltuva:13d}
that the excitation of the core may be an important reaction mechanism and needs to be 
taken into account. There are several formulations of rigorous three-body  scattering 
equations including core excitation (CeX) \cite{mukhamedzhanov:12a,deltuva:13d}, however,
numerical results were obtained only in Ref.~\cite{deltuva:13d}.
As dynamic input these calculations use nucleon-nucleon ($NN$) and nucleon-core ($NA$)
interactions. There is a number of standard parametrizations for $NA$ optical potentials (OP)
without CeX. The CeX is usually included by deforming these standard potentials to allow the
coupling between ground ($A$) and excited ($A^*$) states of the core.
However, this way the additional $NA^*$ component in the two-body scattering equation 
together with the potential deformation distort the elastic $NA$ amplitude that deviates from the 
original elastic amplitude calculated using standard potential without CeX. 
The latter is usually fitted to the $A(N,N)A$ elastic experimental data, 
thus, the description of the data gets destroyed when
including the CeX via the simple deformation of the potential.
Furthermore, if the $NA-NA^*$ coupling potential was fitted to the inelastic 
$A(N,N')A^*$ data using the distorted-wave Born approximation (DWBA) which is the usual case,
the fit to $A(N,N')A^*$ data is lost as well. Thus, an additional adjustment of the OP
parameters needs to be performed to restore the description of $NA$ elastic and inelastic
scattering data. This implies a new fit procedure for each nucleus at each energy, 
which may be quite tedious.
On the other hand, one may take the advantage of already existing fits to the experimental
data obtained using DWBA and standard OP. For this a $NA$ potential with CeX is needed that, 
when inserted into the exact two-body multichannel Lippmann-Schwinger equation, yields
the standard OP results for the $A(N,N)A$  elastic scattering and DWBA results for
the $A(N,N')A^*$ reaction. In the present work I propose a method for achieving this goal,
i.e., I construct a $NA$ potential with CeX that at a given energy exactly reproduces
the standard OP and DWBA amplitudes for $A(N,N)A$ and  $A(N,N')A^*$ scattering processes,
respectively. Thus, the desired consistency with the $NA$ experimental data is ensured.
In case there are no data available, this approach still allows for a more precise evaluation
of the CeX effect, not affected by the OP mismatch.
In three-body systems I aim to study the changes in observables caused by this improvement
of the OP. I therefore reanalyze the reactions calculated in  Ref.~\cite{deltuva:13d} using
the simple deformation of the OP as well as present several new cases.

In Sec.~\ref{sec:2} I derive the $NA$ potential with CeX, and in Sec.~\ref{sec:3}
I recall the three-body scattering theory. Example results for 
$d+\A{10}{Be}$, $p+\A{11}{Be}$, and $d+\A{24}{Mg}$ reactions are presented in 
Sec.~\ref{sec:res}, while the summary is given in  Sec.~\ref{sec:sum}.

\section{Subtraction method for nucleon-core potential \label{sec:2}}

\begin{figure}[!]
\begin{center}
\includegraphics[scale=0.44]{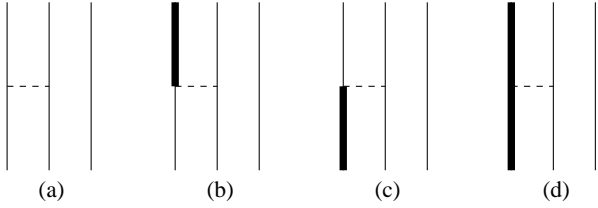}
\end{center}
\caption{\label{fig:vxg} 
Nucleon-core potential in the three-body Hilbert space. 
The particles are represented by vertical lines, the core in its excited state
is distinguished by thick lines, the potential by horizontal dashed lines.
The diagrams (a), (b), (c), and (d) correspond to
$V_{gg}$, $V_{xg}$, $ V_{gx}$, and  $V_{xx}$.}
\end{figure}

I employ  the extended Hilbert space 
$\clh_g \oplus \clh_x$ where the two sectors correspond to the core being in
its ground (g) or excited (x) state \cite{deltuva:13d}.
I aim to construct  the $NA$ potential coupling the  two sectors and denote its components
by $V_{gg}$, $V_{xx}$, and  $V_{xg} =  V_{gx}^T$; they are represented 
graphically in Fig.~\ref{fig:vxg}. The  
coupled-channel Lippmann-Schwinger equation 
\begin{equation}  \label{eq:T}
\begin{pmatrix} T_{gg} \, T_{gx} \\ T_{xg} \, T_{xx} \end{pmatrix}
= \begin{pmatrix} V_{gg}  \, V_{gx} \\ V_{xg}  \, V_{xx} \end{pmatrix}
+ \begin{pmatrix} V_{gg}  \, V_{gx} \\ V_{xg} \,  V_{xx} \end{pmatrix} G_0
\begin{pmatrix} T_{gg}  \, T_{gx} \\ T_{xg}  \, T_{xx} \end{pmatrix}
\end{equation}
yields the respective components $T_{ji}$ of the  $NA$ transition matrix at the available 
energy $E$  with $G_0 = (E+i0 -H_0)^{-1}$ being the free resolvent.
I emphasize that the extended free Hamiltonian $H_0$
besides the kinetic energy operator contains also the internal core Hamiltonian whose contribution
vanishes for $\clh_g$ but is equal to  the core excitation energy $(m_{A^*}-m_A)$ for
$\clh_x$.

At energy $E=E_s$ the component $T_{gg}$ describing the  $NA$ elastic scattering  
is demanded to reproduce the transition matrix
\begin{equation}  \label{eq:tg}
t_g = v_g + v_g G_0 t_g
\end{equation}
obtained with the  $NA$ potential $v_g$ not including CeX, usually taken from one of the standard
parametrizations.
There is no such demand for $T_{xx}$
due to the lack of the experimental data for the  $NA^*$ elastic scattering. The  respective 
transition matrix for the core in its excited state but without coupling to the ground state 
\begin{equation}  \label{eq:tx}
t_x = v_x + v_x G_0 t_x
\end{equation}
is obtained with the $NA^*$ potential $v_x$ that is not constrained by the data but is usually taken 
from the same standard OP parametrization as $v_g$.

Furthermore, at  $E=E_s$  I also require $T_{xg}$ to reproduce the DWBA amplitude
\begin{equation}  \label{eq:dwba}
T_{xg}^{\rm DWBA} = (1+t_x G_0) V_{xg}^{\rm DWBA} (1+G_0 t_g).
\end{equation}
The potential $V_{xg}^{\rm DWBA}$ coupling the two Hilbert sectors $\clh_g$ and $\clh_x$
usually is obtained by deforming the central part of  $v_g$. 
In the  rotational model \cite{nunes:96a} one assumes  that the
core has a permanent quadrupole deformation and replaces in  $v_g$ the nuclear radius
$R_0$ by  $R = R_0(1+\beta_2 Y_{20}(\hat{\xi}))$, where  $\beta_2$
is the quadrupole deformation parameter and  $\hat{\xi}$ describes 
the internal core degrees of freedom  in the body-fixed frame.
If the central part of $v_g$ is a function of $(r-R_0)$, e.g., the
Woods-Saxon function,  $V_{xg}^{\rm DWBA}$ is given by 
$\mathcal{P}_x v_{g}(r-\delta_2Y_{20}(\hat{\xi})) \mathcal{P}_g$ where
$\mathcal{P}_i$ are the respective projectors and 
 $\delta_2 = \beta_2 R_0$ is the deformation length. 

\begin{figure}[!]
\begin{center}
\includegraphics[scale=0.5]{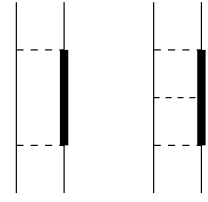}
\end{center}
\caption{\label{fig:vgg} 
Diagrammatic representation of the contributions at the two lowest orders 
  subtracted from the nucleon-core potential.}
\end{figure}

To fulfill the above demands, $T_{xg}$ in Eq.~(\ref{eq:T}) is resolved  as 
\begin{equation}  \label{eq:Txg}
 T_{xg} = (1 -  V_{xx} G_0)^{-1}  V_{xg} (1 + G_0 T_{gg})
\end{equation}
and  used in this for the $T_{gg}$ component, leading to
\begin{equation}  \label{eq:Tgg}
T_{gg} = V_{gg} (1 + G_0 T_{gg}) +  V_{gx} G_0 (1 -  V_{xx} G_0)^{-1}  V_{xg} (1 + G_0 T_{gg}).
\end{equation}
Furthermore, I make use of the identity
\begin{equation}  \label{eq:txv}
(1 -  v_{x} G_0)^{-1} = 1 + t_x G_0
\end{equation}
obtained from  Eq.~(\ref{eq:tx}). By comparing Eqs.~(\ref{eq:Tgg}) and (\ref{eq:Txg})
with (\ref{eq:tg}) and (\ref{eq:dwba}) one observes that the conditions
$T_{gg} = t_g$ and $T_{xg} = T_{xg}^{\rm DWBA}$ are satisfied at  $E=E_s$  by choosing
\begin{subequations}  \label{eq:AGS}   
\begin{align}  
V_{gg} = {}& v_g -  V_{gx} g_0^s (1 +  t_{x} g_0^s)  V_{xg}, \\
V_{xg} = {}& V_{xg}^{\rm DWBA} , \\
V_{xx} = {}& v_x,
\end{align}
\end{subequations}
with $g_0^s = (E_s+i0 -H_0)^{-1}$.
Thus, the essential idea of the present method is subtracting from the elastic amplitude
the contributions that are explicitly generated in the scattering equations
by the coupling to the core excited state.
For this reason it is called the subtraction method.
The subtracted contributions at the two lowest orders are diagrammatically represented in
Fig.~\ref{fig:vgg}. These contributions are nonlocal in the coordinate space but this has no 
disadvantage when the calculations are performed in the momentum space.

Note that a similar subtraction method has been used in the past to readjust the purely
nucleonic part of the two-nucleon potential with explicit $\Delta$-isobar excitation
 \cite{hajduk:83a}, thereby improving the fit of the elastic two-nucleon scattering data.
The present method is more general in the sense that it also fixes
the inelastic amplitude.

\section{Three-particle scattering equations \label{sec:3}}

Exact three-body scattering equations of Faddeev or 
Alt-Grassberger-Sandhas (AGS) type 
including the CeX have been discussed in Refs.~\cite{mukhamedzhanov:12a,deltuva:13d};
their practical solution was implemented in  Ref.~\cite{deltuva:13d}. 
In the extended Hilbert space they acquire the standard form of the
AGS integral equations for three-body transition operators
\begin{equation}  \label{eq:Uba}
U_{\beta \alpha}  = \bar{\delta}_{\beta\alpha} \, G^{-1}_{0}  +
\sum_{\gamma=1}^3   \bar{\delta}_{\beta \gamma} \, T_{\gamma} 
\, G_{0} U_{\gamma \alpha}
\end{equation}
that couple the two sectors  $\clh_g$ and $\clh_x$ much like the two-body
transition operators $T_{\gamma}$ in Eq.~(\ref{eq:T}).
The subscripts $\alpha, \beta, \gamma$ label the spectator particles 
(interacting pairs in the odd-man-out notation)
for operator components, while $ \bar{\delta}_{\beta\alpha} = 1 - \delta_{\beta\alpha}$.
Following the developments of Ref.~\cite{deltuva:13d}, I solve 
Eq.~(\ref{eq:Uba}) numerically in the momentum-space partial-wave framework, including the
proton-core Coulomb force via the  screening and renormalization method
\cite{taylor:74a,alt:80a,deltuva:05a}. 
The scattering amplitudes are given by the on-shell matrix elements of the 
 transition operators $U_{\beta \alpha}$ calculated between initial and final
channel states.

\section{Results \label{sec:res}}

In this section I present results for two-cluster elastic, inelastic, and transfer reactions
initiated  by $p+\A{10}{Be}$,  $d+\A{10}{Be}$,
 $p+\A{11}{Be}$, and  $d+\A{24}{Mg}$ collisions. For both $\A{10}{Be}$ and $\A{24}{Mg}$
cores the ground (first excited) states have spin and parity $0^+$ ($2^+$),
while the  respective excitation energies are 3.368  and 1.369 MeV.
The main goal is to evaluate the changes in the observables due to the use of the
improved nucleon-core potential including the CeX with subtraction.
I present three types of calculations. All of them use a realistic
CD Bonn potential \cite{machleidt:01a} for the $np$ pair, but differ in $NA$
interactions:
(i) single-particle (SP) model for the core, i.e., neglecting CeX in $NA$ potentials,
labeled as SP in the following;
(ii) including CeX via simple deformation of the SP potential without subtraction and 
readjustment, labeled CX(no subtr) in the following; and 
(iii) including CeX using improved $NA$ potentials with subtraction and proper
readjustment as described in Sec.~\ref{sec:2}, labeled CX  in the following.
Thus, the difference between CX and SP will yield the the CeX effect, while
the difference between CX and  CX (no subtr) will evaluate the 
importance of the potential improvement.
The calculations SP and CX(no subtr), both based on the Chapel Hill 89 (CH89)
parametrization \cite{CH89},  are 
the same as in Ref.~\cite{deltuva:13d}, but CX(no subtr) was labeled simply CX 
in Ref.~\cite{deltuva:13d}. I therefore only have to describe the calculations
CX with subtraction. For the $nA$ pair in the partial waves with $(An)$ bound states
I take over the real binding potentials from  Ref.~\cite{deltuva:13d} since
they are already adjusted to experimental binding energies and need no subtraction.
Depending on the partial wave, these potentials may
support deeply bound states $|b_0 \rangle$ that are Pauli forbidden and therefore have to
be projected out;  this is achieved
by adding a separable term  $|b_0 \rangle \Gamma \langle |b_0 |$ 
with $\Gamma \ge 1$ GeV to the local  $nA$ potential 
and thereby  moving the Pauli forbidden state  $|b_0 \rangle$  to a large positive energy
\cite{schellingerhout:93a}.
For  $nA$ in other partial waves and for $pA$  in all partial waves
I take the potentials including CeX with subtraction; they are
 derived from the CH89 parametrization unless explicitly stated otherwise. 
The subtraction energy $E_s$ coincides with the energy at which the OP is taken, i.e.,
half of the deuteron energy $E_d$ for $nA$ but proton energy $E_p$ for $pA$.
However, if the $nA$ potential is complex in all partial waves  as in the case of the present
$d+\A{24}{Mg}$  calculations, the $p + (An)$ channel is absent and therefore
the $pA$ potential is taken at $E_d/2$ as well.
These choices are the same as in Ref.~\cite{deltuva:13d}.
The employed quadrupole deformation parameters
for $N$-$\A{10}{Be}$ interactions are $\beta_2 = 0.67$ and $\delta_2 = 1.664$ fm
as in Ref.~\cite{deltuva:13d},
while for $N$-$\A{24}{Mg}$ I use  $\beta_2 = 0.5$ and $\delta_2 = 1.69$ fm.
In a few selected cases I present CX results derived from Watson
\cite{watson} and Koning-Delaroche \cite{koning} parametrizations; those calculations
use the $\delta_2$ values listed above and are labeled as CX(W) and CX(KD), respectively.

\begin{figure}[!]
\begin{center}
\includegraphics[scale=\scl]{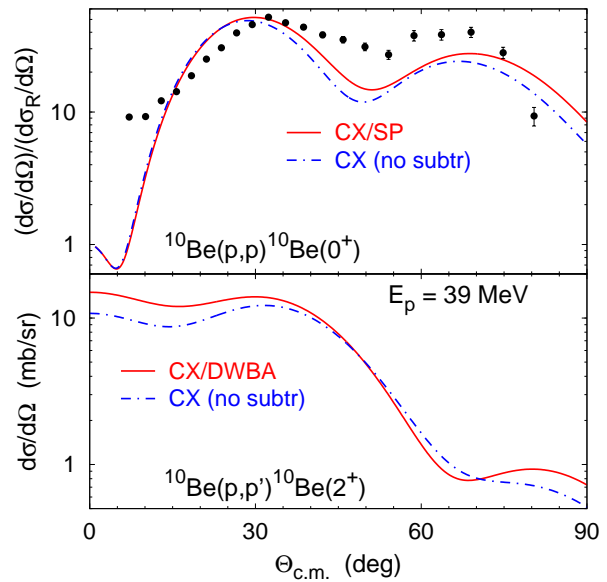}
\end{center}
\caption{\label{fig:p10Be}  (Color online)
Differential cross section for $\A{10}{Be}(p,p)\A{10}{Be}$
and $\A{10}{Be}(p,p')\A{10}{Be}^*$  reactions  at $E_p = 39$ MeV.
 Results including the CeX with and without
subtraction, i.e., CX and CX(no subtr), are given by solid and dashed-dotted curves, respectively.
CX predictions  coincide with SP results neglecting the CeX
for the elastic scattering and  with DWBA results for the inelastic reaction.
The elastic experimental data are from Ref.~\cite{lapoux:07}.}
\end{figure}

First in Fig.~\ref{fig:p10Be} I demonstrate the importance of the proper potential readjustment  
by comparing the CX and CX(no subtr) results in the two-body system.
As an example I show the differential cross section $d\sigma/d\Omega$ for elastic and inelastic 
$p+\A{10}{Be}$ scattering at $E_p = 39$ MeV proton energy
as a function of the center-of-mass (c.m.) scattering angle $\Theta_{\mathrm{c.m.}}$.
Here and in the following the elastic differential cross section is given as a 
ratio to the Rutherford cross section  $d\sigma_R/d\Omega$.
The differences between the CX and CX(no subtr)
calculations are of moderate size for
elastic scattering but become more significant for the 
 $\A{10}{Be}(p,p')\A{10}{Be}^*$ reaction, reaching nearly 40\% at forward angles.
I remind that by construction the CX and SP results coincide 
 for the elastic  scattering while the CX and DWBA  results coincide 
for the inelastic  scattering.

\begin{figure}[!]
\begin{center}
\includegraphics[scale=\scl]{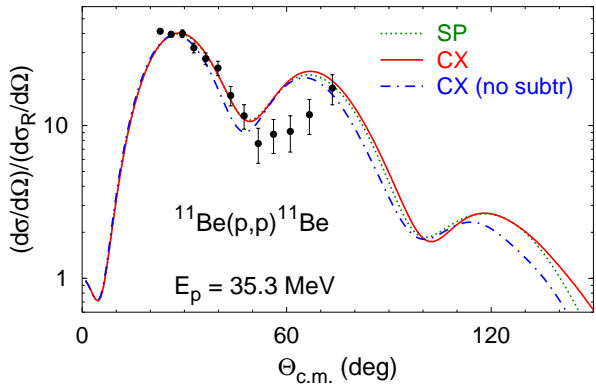}
\end{center}
\caption{\label{fig:p11Be}  (Color online)
Differential cross section for 
$p+\A{11}{Be}$  elastic scattering  at $E_p = 35.3$ MeV.
 Results of SP, CX, and CX(no subtr) potential models are given
by dotted, solid, and dashed-dotted curves, respectively.
The experimental data are from Ref.~\cite{lapoux:07}.}
\end{figure}

Observables for $p+\A{10}{Be}$ and $p+\A{11}{Be}$ elastic scattering are quite
strongly correlated. Consistently with this fact the SP and CX models, yielding
identical results for $\A{10}{Be}(p,p)\A{10}{Be}$, agree quite well also for
 $\A{11}{Be}(p,p)\A{11}{Be}$ at $E_p = 35.3$ MeV as shown in Fig.~\ref{fig:p11Be},
indicating that the CeX effect on the $p+\A{11}{Be}$ elastic cross section is very small.
However, the $p$-$\A{10}{Be}$ potential, that does not properly describes elastic
two-body data as in the case of CX(no subtr), overestimates the CeX effect.

\begin{figure}[!]
\begin{center}
\includegraphics[scale=0.56]{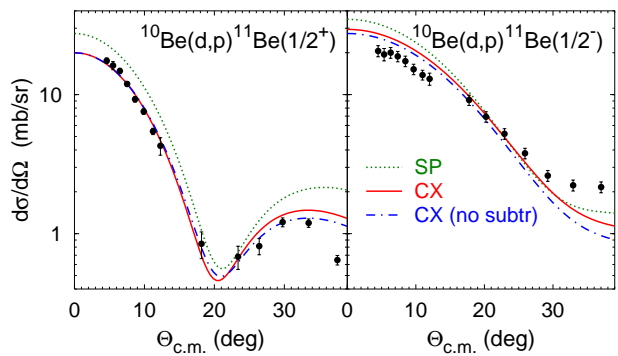}
\end{center}
\caption{\label{fig:d10Be}  (Color online)
Differential cross section for transfer reactions 
$\A{10}{Be}(d,p)\A{11}{Be}$  at $E_d =  21.4$ MeV
leading to the ground ($\frac12^+$) and excited ($\frac12^-$)
 states of $\A{11}{Be}$.
Curves are as in Fig.~\ref{fig:p11Be} and 
experimental data are from Ref.~\cite{dBe12-21}.}
\end{figure}

\begin{figure}[!]
\begin{center}
\includegraphics[scale=0.56]{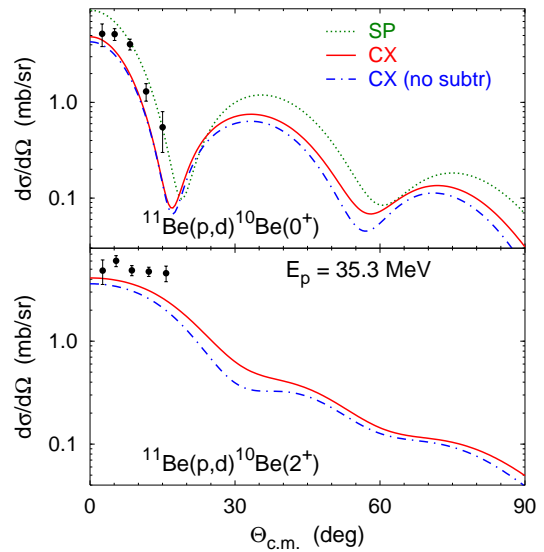}
\end{center}
\caption{\label{fig:p11Bed}  (Color online)
Differential cross section for
$\A{11}{Be}(p,d)\A{10}{Be}$  transfer reactions at $E_p = 35.3$ MeV
leading to the ground ($0^+$) and excited ($2^+$) states of $\A{10}{Be}$.
Curves are as in Fig.~\ref{fig:p11Be} and 
experimental data are from Ref.~\cite{winfield:01}.}
\end{figure}

\begin{figure}[!]
\begin{center}
\includegraphics[scale=0.56]{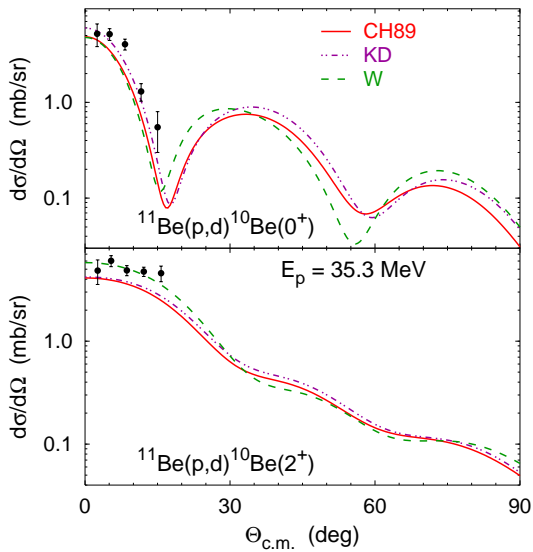}
\end{center}
\caption{\label{fig:p11Bed-v}  (Color online)
Differential cross section for  
$\A{11}{Be}(p,d)\A{10}{Be}$ transfer reactions  at $E_p = 35.3$ MeV
leading to the ground ($0^+$) and excited ($2^+$) states of $\A{10}{Be}$.
CX predictions based on CH89 (solid curves), Koning-Delaroche (double-dotted-dashed curves), 
and Watson (dashed curves) potentials are compared with the
experimental data from Ref.~\cite{winfield:01}.}
\end{figure}

Next I show in Fig.~\ref{fig:d10Be} the differential cross sections for 
the $\A{10}{Be}(d,p)\A{11}{Be}$ transfer reactions at $E_d = 21.4$ MeV deuteron energy.
Quite surprisingly, the improvement of the $N$-$\A{10}{Be}$  potentials has a very small
effect on the transfer to the $\A{11}{Be}$ ground state $\frac12^+$.
For the transfer to the $\A{11}{Be}$ excited state $\frac12^-$ the effect is of moderate
size but not beneficial. Thus, the $\A{10}{Be}(d,p)\A{11}{Be}^*$ reaction still remains 
 an unresolved problem and calls for a more sophisticated model.
In Fig.~\ref{fig:p11Bed} I present results for
$\A{11}{Be}(p,d)\A{10}{Be}$ transfer reactions at $E_p = 35.3$ MeV;
it corresponds to  $\A{10}{Be}(d,p)\A{11}{Be}$ at $E_d = 40.3$ MeV.
At this higher energy the importance of the proper potential adjustment is more visible,
especially for the reaction leading to the $2^+$ excited state of the $\A{10}{Be}$ core.
The latter finding is not unexpected given the results
for  $p+\A{10}{Be}$ inelastic scattering in Fig.~\ref{fig:p10Be}.
For both ground and excited $\A{10}{Be}$ states the description of the experimental
data is improved, although some discrepancies still remain.
In fact, these predictions are quite sensitive to the choice of the optical potential
from which the CX model is derived. I illustrate this finding in Fig.~\ref{fig:p11Bed-v} by
comparing the CX results for the $\A{11}{Be}(p,d)\A{10}{Be}$ differential cross sections based on
CH89, Watson, and Koning-Delaroche OP parametrizations.
I admit that no one of them reproduces the experimental data 
for both reactions simultaneously: transfer to the $\A{10}{Be}$ ground state $0^+$ is 
best described by CX(KD) while to the excited state $2^+$ by CX(W).

\begin{figure}[!]
\begin{center}
\includegraphics[scale=0.55]{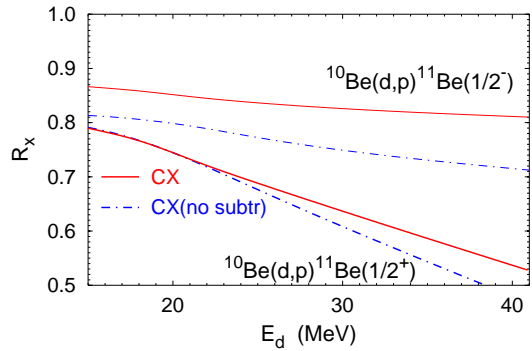}
\end{center}
\caption{\label{fig:r}  (Color online)
Differential cross sections ratios $R_{x}$ for 
$\A{10}{Be}(d,p)\A{11}{Be}$ transfer reactions  
leading to ground (thick curves) and excited (thin curves) states
of $\A{11}{Be}$. CX and CX(no subtr) predictions are given by
solid and dashed-dotted curves, respectively.}
\end{figure}

In Ref.~\cite{deltuva:13d} it was demonstrated that the CeX effect in transfer reactions is,
in general, much more complicated than just a simple rescaling of 
the SP differential cross section by the respective spectroscopic factor (SF).
This is in contrast with naive assumptions often used for the SF extraction by comparing
DWBA-type SP calculations and experimental data.
In Fig.~\ref{fig:r} I confirm these findings of Ref.~\cite{deltuva:13d} also when
using an improved potential with CeX. For $x$ being either CX or CX(no subtr) 
I show the ratios $R_{x} = (d\sigma/d\Omega)_{x}/(d\sigma/d\Omega)_{\rm SP}$ 
in $\A{10}{Be}(d,p)\A{11}{Be}$  reactions at
$\Theta_{\mathrm{c.m.}}=0^\circ$  as functions of the deuteron energy $E_d$.
The deviation of $R_{x}$ from the SF, which equals to 0.855 (0.786) for the 
$\A{11}{Be}$ ground (excited) state, becomes most evident for the transfer
to the $\A{11}{Be}$ ground state at higher  $E_d$.
Both CX and CX(no subtr) models show qualitatively the same behavior.
The difference between them increases with increasing $E_d$ and thereby indicates the 
importance of the proper fit to the two-body data, but does not alter the general conclusion.

\begin{figure}[!]
\begin{center}
\includegraphics[scale=0.58]{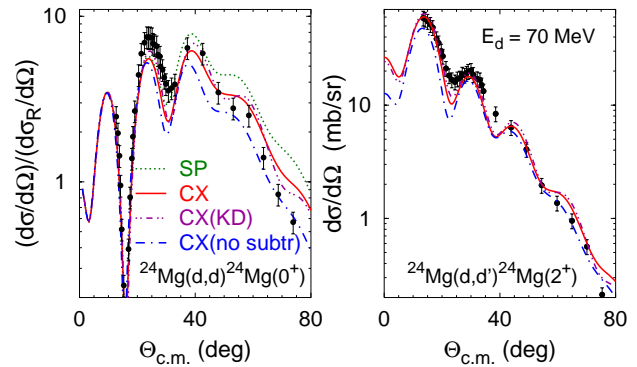}
\end{center}
\caption{\label{fig:mg}  (Color online)
Differential cross section for  $d + \A{24}{Mg}$ elastic (left)
and inelastic (right) scattering at $E_d = 70$ MeV.
Results of SP, CX, and CX(no subtr) models based on the CH89 potential are given
by dotted, solid, and dashed-dotted curves, respectively,
while CX results based on the Koning-Delaroche potential
are given by double-dotted-dashed curves. 
The experimental data are from Ref.~\cite{mg:d}.}
\end{figure}

Finally I consider the $d+\A{24}{Mg}$ scattering.
The quadrupole deformation $\beta_2$ for $\A{24}{Mg}$ determined in different
experiments varies between 0.4 and 0.6 (see Ref.~\cite{mg:d} for overview).
In Ref.~\cite{deltuva:13d} the CX(no subtr) calculations  were performed
with $\beta_2 = 0.4$ and 0.6, with the lower value favored by the 
$\A{24}{Mg}(d,d)\A{24}{Mg}$ data and the upper value favored by the  $\A{24}{Mg}(d,d')\A{24}{Mg}^*$ data,
but no one of them was able to reproduce 
the data for both reactions simultaneously.
Here I present in Fig.~\ref{fig:mg} the results for $d+\A{24}{Mg}$ elastic and inelastic
differential cross section at $E_d = 70$ MeV obtained with $\beta_2 = 0.5$.
The CX model yields the best description of the experimental data and fits reasonably
well both elastic and inelastic cross sections, while the CX(no subtr) model
overestimates the CeX effect for the $\A{24}{Mg}(d,d)\A{24}{Mg}$ 
but underestimates  for the $\A{24}{Mg}(d,d')\A{24}{Mg}^*$ reaction.
Changing the starting potential from CH89 to Koning-Delaroche
has only a small effect as demonstrated by the CX(KD) results.

\section{Summary \label{sec:sum}}

I considered elastic, inelastic, and transfer reactions in three-body nuclear systems 
consisting of a neutron, a proton, and a core.
I explicitly included the core excitation in the exact scattering equations and
 solved them in the momentum-space framework. The calculational technique was taken
over from Ref.~\cite{deltuva:13d} but the dynamic input was significantly improved.

The technical objective of this work was to develop
 a method for constructing the nucleon-core potential 
that couples ground and excited states of the core and in the coupled-channel two-body 
Lippmann-Schwinger equation
reproduces the predictions of the given standard optical potential
for the elastic scattering as well as the DWBA predictions for the inelastic reaction.
The essence of the method is subtracting  from the original single-channel
potential the explicit core excitation contributions,
thereby avoiding the double counting.
In the usual case where a coupled-channel potential properly fitted to the experimental
data is not available but the nucleus deformation parameters are determined
in the DWBA, the proposed method yields the coupled-channel potential
including the core excitation and consistent with the data for
both elastic and inelastic scattering. This is an important improvement
over the simple deformation of the potential used previously \cite{deltuva:13d}
that destroys the fit to the experimental data. 
For the $p + \A{10}{Be}$ example it is demonstrated that this deviation may be significant,
especially for the inelastic $(p,p')$ reaction. 

The physics objective of the present work was the evaluation of  changes in 
three-body observables due to the improvement of employed optical potentials.
In three-body reactions involving $\A{10}{Be}$ and $\A{24}{Mg}$ cores
the most important changes  were found in elastic $(p,p)$ and
$(d,d)$, inelastic $(d,d')$, and transfer  $(p,d)$  reactions leading
to the core in its excited state.
Transfer reactions $(d,p)$  were affected less, at least at lower energies. 
In $p+\A{11}{Be}$ elastic scattering the core excitation effect turns out to be very small.
Compared to results of Ref.~\cite{deltuva:13d},
the improved nucleon-core potential leads to larger cross sections in most cases.
This  increases slightly the discrepancy between predictions and data for the
$\A{10}{Be}(d,p)\A{11}{Be}^*$ transfer reaction but improves the agreement for the
$\A{11}{Be}(p,d)\A{10}{Be}^*$ reaction. In the latter case I also found a significant sensitivity
to the parametrization of the standard optical potential that serves as a starting point
in the calculations. 
Concerning the $d+\A{24}{Mg}$ scattering, the improved calculations
are able to describe elastic and inelastic cross sections simultaneously with the same value 
of the deformation parameter $\beta_2=0.5$, in contrast to Ref.~\cite{deltuva:13d}.
Finally, I qualitatively confirmed 
the findings of  Ref.~\cite{deltuva:13d} that the effect of the core excitation
in transfer reactions cannot be simply related to the respective spectroscopic factor.

\vspace{1mm}



\end{document}